\documentclass{goose-article}

\hypersetup{pdfauthor={%
  T.W.J. de Geus%
}}
\header{%
  T.W.J.~de~Geus, F.~Maresca, R.H.J.~Peerlings, M.G.D.~Geers\\%
  Mechanics of Materials, 2016, 101:147-159, %
  \doi{10.1016/j.mechmat.2016.07.014}, \eprint{1603.05847}%
}

\title{%
  Microscopic plasticity and damage in two-phase steels: on the competing role of crystallography and phase contrast%
}
\author[1,2]{T.W.J.~de~Geus}
\author[1,2]{F.~Maresca}
\author[2]{R.H.J.~Peerlings}
\author[2]{M.G.D.~Geers}
\affil[1]{%
  Materials innovation institute (M2i), Delft, The Netherlands%
}
\affil[2]{%
  Department of Mechanical Engineering, Eindhoven University of Technology, Eindhoven, The Netherlands%
}
\contact{%
  $^*$Corresponding author: \href{mailto:t.w.j.d.geus@tue.nl}{t.w.j.d.geus@tue.nl} -- \href{mailto:tom@geus.me}{tom@geus.me} %
}

\begin{document}

\maketitle

\begin{abstract}%
This paper unravels micromechanical aspects of metallic materials whose microstructure comprises grains of two or more phases. The local plastic response is determined by (i) the relative misorientation of the slip systems of individual grains, and (ii) the different mechanical properties of the phases. The relative importance of these two mechanisms at the meso-scale is unclear: is the plastic response dominated by the grain's anisotropy, or is this effect overwhelmed by the mechanical contrast between the two phases? The answer impacts the modeling of such a material at the meso-scale, but also gives insights in the resulting fracture mechanisms at that length-scale. Until now, this question has been addressed only for particular crystallographies and mechanical properties. In contrast, this paper studies the issue systematically using a large set of phase distributions, crystallographies, and material parameters. It is found that the macroscopic and the mesoscopic (grain-averaged) plastic response of the two extreme modeling choices (crystal plasticity or isotropic plasticity) converge with increasing phase contrast. The effect of the crystallography is completely overwhelmed by the phase contrast when the yield stress of the hard phase is a factor of four higher compared to the soft phase. When this ratio is lower than two, its influence may not be neglected. However, even in this regime, fracture initiation is controlled by the local arrangement of the phases. The latter is quantified in this paper through the average arrangement of the phases around fracture initiation sites.
\end{abstract}

\keywords{crystal plasticity; two-phase metals; dual-phase steel micromechanics; ductile damage}

\section{Introduction}

This paper studies the plastic response of two-phase metals that from an engineering perspective combine strength with ductility. The microstructure typically consists of two phases, a soft and a hard phase. We study the competition between plasticity due to the misalignment of the slip systems of the grains, and that due to the difference in slip resistance between the two phases. We aim to determine their relative influence on the plastic response at the meso-scale level (which has a grain-averaged resolution); showing that at that level the underlying physics may potentially be considered in an averaged sense. From a modeling point of view this implies that an isotropic plasticity model may provide sufficiently accurate predictions. Such a model is substantially simpler and computationally less expensive than the more realistic crystal plasticity model. Earlier works in this direction mainly focused on a single set of parameters or microstructures. In contrast, we consider a statistically relevant set of random microstructures for a wide range of parameters, thereby reconciling apparently contradictory conclusions in different regimes.

In accordance with the considered grain-averaged quantities, an idealized microstructural morphology comprising equi-sized square grains is used. This offers the advantage of well controlled variations. Even more important is that the computations are relatively inexpensive as few finite elements suffice to discretize the microstructure accurately. For more targeted studies, involving only one or a few microstructures, such a simplification is not strictly needed and more realistic morphologies, in three dimensions, may be used \citep{Zhao2008,Wong2015,Proudhon2016,Vachhani2015,Delannay2006,Melchior2006}.

Materials that belong to the category of interest are metal-matrix composites and advanced high strength steels. The metal-matrix composites often have a (soft) aluminum matrix reinforced by (hard) silicon-carbide particles. It was recognized that in that case the response is dominated mostly by the strong contrast in properties of the two constituents. The crystallography of the matrix was found to matter only at the final stages of deformation where fracture initiates \cite{Aghababaei2011,Needleman1993,McHugh1993,Nugent2000}. Advanced high strength steels have a considerably more complicated, fully metallic, microstructure, whereby the main mechanical contributions originate from the (soft) ferrite and the (hard) martensite phase. These steels display many phenomena which cannot be trivially explained by the theory and models that have been proposed for the, more simple, metal-matrix composites. Existing numerical studies were extended with more realistic microstructures \cite[e.g.][]{Choi2009,Sun2009a,Kumar2006} but also with more realistic constitutive models, often involving crystal plasticity \cite[e.g.][]{Choi2013,Tjahjanto2006,Roters2010,Temizer2016,Maresca2016}.

For single phase materials, the effects of crystallography on the plastic and damage response are significant, as the misalignment of grains introduces mechanical contrast \cite{Asaro1985}. The natural question that arises is to what extent the phase contrast in two-phase materials dominates over this \textit{misalignment contrast}. It is known that for the case of rigid particles embedded in an elasto-plastic matrix, the micromechanical response is mostly controlled by the interaction of the particles and the matrix phase while the crystallography is less important \cite{Needleman1993,McHugh1993,Nugent2000}. Although most studies are obviously numerical, \citet{Nugent2000} present an experiment in which rigid fiber is embedded in an analog matrix material (silver chloride). Local measurements using transmission electron microscopy were compared to isotropic elasto-plastic simulations. A satisfactory accuracy of the model predictions compared to the experiments was found except for observations with sub-grain resolution. For dual-phase steels, wherein also the hard phase deforms plastically, the crystallography appears to play an important role that may not be neglected \cite{Choi2013,Diehl2015}.

These studies were however mostly focused on the sub-grain response based on a specific set of crystallographies and with parameters (mechanical and crystallographic / morphological) reflecting extreme cases. This paper considers a much wider range of parameters to identify where the two extreme regimes end. The analysis is extended to the initiation of fracture for which the outcome of the competition between crystallography and phase contrast is yet undetermined. In particular for this case it is imperative to consider a large set of microstructures.

This paper is structured as follows. The microstructural model and the crystal and isotropic plasticity models are introduced in Sections~\ref{sec:micro} and~\ref{sec:const}. The macroscopic and microscopic elasto-plastic responses for these two models are compared in Section~\ref{sec:response} for a wide range of hard phase volume fractions and phase contrasts. This comparison is extended to the initiation of ductile fracture in Section~\ref{sec:damage}. The paper ends with concluding remarks in Section~\ref{sec:conclusion}.

\section*{Nomenclature}

\begin{tabular}{llll}
$\langle \ldots \rangle$                      & ensemble average
\\
$\bar{a}$                                     & volume average
\\
$\bm{A}$                                      & second order tensor
\\
$\mathbb{A}$                                  & fourth order tensor
\\
$\mathbb{C} \, = \bm{A} \otimes \bm{B}$       & dyadic tensor product
\\
$\bm{C} = \bm{A} \cdot \bm{B}$                & single tensor contraction
\\
$c \;\, = \bm{A} : \bm{B} = A_{ij} B_{ji}$    & double tensor contraction
\\
\end{tabular}

\section{Microstructural model}
\label{sec:micro}

\subsection{Microstructure and spatial discretization}

\begin{figure}[htp]
  \centering
  \includegraphics[width=.75\textwidth]{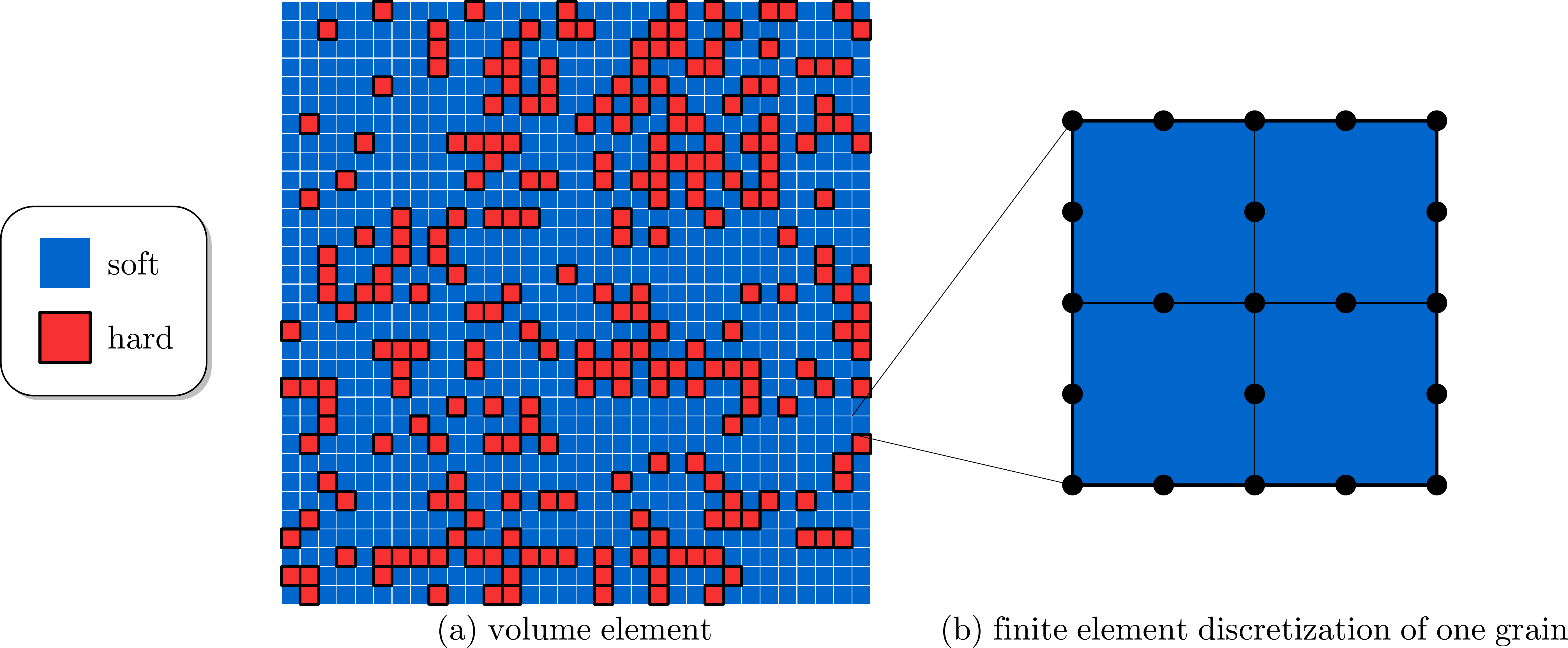}
  \caption{(a) A typical microstructure from the ensemble with a hard phase volume fraction $\varphi^\mathrm{hard} = 0.25$, whereby the grains of the two phases are shown in blue and red. (b) The numerical discretization within a single grain.}
  \label{fig:typical_I_2D}
\end{figure}

The microstructure is represented by an ensemble of $256$ idealized periodic unit-cells, each made up of $32 \times 32$ equi-sized \textit{grains} in which the phases are randomly distributed. An example unit-cell is shown in Figure~\ref{fig:typical_I_2D}(a). Because we are interested in the \textit{grain-averaged} (tensorial) quantities, the morphological idealization enables a relatively coarse numerical discretization, as discussed below.

Not the individual unit-cells but the ensemble as a whole is considered to be representative for the material. Accordingly, the hard phase volume fraction is set for the ensemble, but fluctuations are allowed between the individual unit-cells. Each grain is randomly assigned the properties of the hard or the soft phase, by comparing a random number in the range $[0,1]$ to a target volume fraction of the hard phase, $\varphi^\mathrm{hard}$. A reference value $\varphi^\mathrm{hard} = 0.25$ is considered, resulting in unit-cells with hard phase volume fractions of $\varphi^\mathrm{hard} = 0.25 \pm 0.04$. The crystallography of the grains is discussed in the next section.

A finite element based solution scheme is used, whereby each grain is discretized using $2 \times 2$ eight node bi-quadratic quadrilateral elements (see Figure~\ref{fig:typical_I_2D}(b)), each of which is numerically integrated using four Gauss points. The quantities of interest are then averaged over the total of 16 Gauss points in each grain. A mesh convergence study has been performed to verify that, using the described discretization, the maximum relative error in terms of the considered quantities is lower than $2\%$ with respect to a much finer discretization of $6 \times 6$ quadratic elements per grain.

\subsection{Macroscopic deformation and periodicity}

The periodicity of the unit-cell is applied using standard periodic boundary conditions. The displacement fluctuations are constrained such that the average deformation gradient tensor equals the prescribed macroscopic deformation gradient $\bar{\bm{F}}$. Plane strain is assumed for the out-of-plane direction.

To concentrate on the local effects in the microstructure, an isochoric macroscopic deformation is applied. This way, any local volumetric deformation --  resulting in a non-zero hydrostatic stress -- can be directly linked to the local microstructural features. Pure shear is used for this purpose, which corresponds to the following logarithmic strain tensor
\begin{equation}
\label{eq:micro:F}
  \bar{\bm{\varepsilon}} =
  \frac{\sqrt{3}}{2} \; \bar{\varepsilon} \;
  \big( \vec{e}_x\vec{e}_x - \vec{e}_y\vec{e}_y \big)
\end{equation}
where $\bar{\varepsilon}$ represents the equivalent logarithmic strain and $\vec{e}_x$ and $\vec{e}_y$ are the Cartesian basis vectors, respectively in horizontal and vertical direction. To introduce sufficient plastic deformation, $\bar{\varepsilon} = 0.2$ is used (applied in $2000$ steps for accurate time integration, see below).

\section{Constitutive models}
\label{sec:const}

\subsection{Introduction}

The response of a standard crystal plasticity model is compared to an isotropic plasticity model. For both models, a rate-dependent formulation is adopted. This choice is classical for crystal plasticity, since it regularizes the slip system activation, while the choice for the isotropic plasticity model is made here for consistency. The parameters are chosen such that the resulting response is close to the rate-independent limit and such that both models coincide in the isotropic limit (i.e.\ an aggregate of grains with identical material properties). The actual values of the yield stress and hardening are therefore different for both constitutive models, even though the same symbols are used below.

Both models are formulated in a finite deformation setting, by assuming the multiplicative split of the deformation gradient, $\bm{F}$, into an elastic part, $\bm{F}_\mathrm{e}$, and plastic part, $\bm{F}_\mathrm{p}$:
\begin{equation} \label{eq:cmod:1}
  \bm{F}
  =
  \bm{F}_\mathrm{e} \cdot \bm{F}_\mathrm{p}
\end{equation}
This split introduces an intermediate configuration distorted by the plastic deformation only. The elastic deformation, as well as rotations, are included in $\bm{F}_\mathrm{e}$.

The plastic part of the velocity gradient is introduced as $\bm{L}_\mathrm{p} := \dot{\bm{F}}_\mathrm{p} \cdot \bm{F}_\mathrm{p}^{-1}$. Different expressions for $\bm{L}_\mathrm{p}$ are chosen for the crystal and the isotropic plasticity model, see below.

It is furthermore useful to introduce the pull-back of the Kirchhoff stress tensor $\bm{\tau}$ to the intermediate configuration:
\begin{equation} \label{eq:cmod:2}
  \bm{S}_\mathrm{i}
  =
  \bm{F}_\mathrm{e}^{-1} \cdot \bm{\tau} \cdot \bm{F}_\mathrm{e}^{-T}
\end{equation}
The elastic constitutive relation for $\bm{S}_\mathrm{i}$ reads:
\begin{equation} \label{eq:cmod:3}
  \bm{S}_\mathrm{i}
  =
  \mathbb{C} : \bm{E}_\mathrm{e}
\end{equation}
for both crystal and isotropic plasticity. Herein, $\mathbb{C}$ is the usual (isotropic) fourth-order elasticity tensor parametrized by Young's modulus $E$ and Poisson's ratio $\nu$. $\bm{E}_\mathrm{e} := \tfrac{1}{2}(\bm{C}_\mathrm{e} - \bm{I})$ is the elastic Green-Lagrange strain, where $\bm{C}_\mathrm{e} := \bm{F}_\mathrm{e}^T \cdot \bm{F}_\mathrm{e}$ is the elastic Cauchy-Green tensor.

\subsection{Crystal plasticity}
\label{sec:const:cp}

Within a standard crystal plasticity formulation \cite{Bronkhorst1992}, the plastic part of the velocity gradient, $\bm{L}_\mathrm{p}$, is the sum of the plastic slip rates, $\dot{\gamma}_{\alpha}$, along the individual slip planes, $\alpha$:
\begin{equation} \label{eq:cmod:4}
  \bm{L}_\mathrm{p}
  =
  \sum_{\alpha=1}^{n_s} \dot{\gamma}_{\alpha} \, \bm{P}_0^{\alpha}
\end{equation}
where $n_s$ is the number of slip systems; the Schmid tensor $\bm{P}_0^{\alpha}$ defines the slip system's orientation through $\bm{P}_0^{\alpha} := \bm{s}_0^{\alpha} \otimes \bm{n}_0^{\alpha}$ where $\bm{s}_0^{\alpha}$ is the slip direction and $\bm{n}_0^{\alpha}$ is the normal to the slip plane in the intermediate configuration. For both the soft and the hard phase, the number of active slip systems $n_s = 12$, all belonging to the $\{110\} \langle 111 \rangle$ slip system family for body centered cubic (bcc) crystals \citep[see][]{Franciosi1983}. The assumption of bcc crystalline symmetry is made \citep{Groger2008,Caillard2010}. Non-Schmid effects are not considered. Throughout this paper, the grain orientations -- i.e.\ the Euler angles -- are selected randomly, such that the aggregate of grains is isotropic \cite[e.g.][]{Yadegari2014}.

The plastic slip rate along each slip system, $\dot{\gamma}_{\alpha}$, is determined via a visco-plastic slip law \cite{Hutchinson1976}
\begin{equation} \label{eq:cmod:5}
  \dot{\gamma}_{\alpha}
  =
  \dot{\gamma}_0
  \left(
    \frac{|\tau_{\alpha}|}{s_{\alpha}}
  \right)^{1/m}
  \mathrm{sign}(\tau_{\alpha})
\end{equation}
where $\dot{\gamma}_0$ is a reference slip rate, $m$ is the strain rate sensitivity, and $\tau_\alpha$ is the resolved shear stress along the $\alpha^{\mathrm{th}}$ slip system:
\begin{equation} \label{eq:cmod:6}
  \tau_{\alpha}
  =
  \big( \bm{S}_\mathrm{i} \cdot \bm{C}_\mathrm{e} \big) : \bm{P}_0^{\alpha}
\end{equation}
The current slip resistance $s_{\alpha}$ is calculated via the following evolution law
\begin{equation} \label{eq:cmod:7}
  \dot{s}_{\alpha}
  =
  \sum_{\beta=1}^{n_s} h_{\alpha \beta} \; |\dot{\gamma}_{\beta}|
\end{equation}
which has an initial value of $s_0$. The hardening is a combination of self-hardening and latent hardening defined through the hardening matrix
\begin{equation} \label{eq:cmod:8}
  h_{\alpha \beta}
  =
  h_0 \left( 1 - \frac{s_{\alpha}}{s_{\infty}} \right)^a
  \big( \delta_{\alpha \beta} + q_n ( 1 - \delta_{\alpha \beta}) \big)
\end{equation}
where $q_n$ is the latent hardening over self-hardening ratio for non-coplanar slip systems and $\delta_{\alpha \beta}$ is the Kronecker delta; $h_0$, $s_{\infty}$ and $a$ are material parameters.

An equivalent plastic strain, $\varepsilon_\mathrm{p}$, is defined for comparison purposes. Its evolution reads
\begin{equation}
\label{eq:const:epdot}
  \dot{\varepsilon}_\mathrm{p} =
  \sqrt{\tfrac{2}{3} \bm{D}_\mathrm{p} : \bm{D}_\mathrm{p}}
\end{equation}
Its initial value is naturally zero.

\subsection{Isotropic plasticity}
\label{sec:const:iso}

For the case of isotropic plasticity, it is assumed that $\bm{L}_\mathrm{p} \equiv \bm{D}_\mathrm{p}$. The following expression of $\bm{D}_\mathrm{p}$ is employed \cite[e.g.][]{Simo1998,Belytschko2009}:
\begin{equation} \label{eq:cmod:9}
  \bm{D}_\mathrm{p}
  =
  \frac{3}{2}
  \frac{\dot{\varepsilon}_\mathrm{p}}{s_\mathrm{eq}} \,
  \bm{S}_\mathrm{i}^{d} \cdot \bm{C}_\mathrm{e}
\end{equation}
where the stress deviator $\bm{S}_\mathrm{i}^{d} = \bm{S}_\mathrm{i} - \tfrac{1}{3} \, ( \bm{S}_\mathrm{i}:\bm{C}_\mathrm{e} ) \, \bm{C}_\mathrm{e}^{-1}$ \cite[cf.][]{Simo1987,Belytschko2009}, and $\dot{\varepsilon}_\mathrm{p}$ is the equivalent plastic strain rate. The equivalent stress $s_\mathrm{eq}$ is defined as follows
\begin{equation} \label{eq:cmod:10}
  s_\mathrm{eq}
  =
  \sqrt{\frac{3}{2}
  ( \bm{S}_\mathrm{i}^{d} \cdot \bm{C}_\mathrm{e} )
  :
  ( \bm{S}_\mathrm{i}^{d} \cdot \bm{C}_\mathrm{e} )}
\end{equation}
The evolution of the plastic strain reads
\begin{equation}
  \dot{\varepsilon}_\mathrm{p} = \dot{\gamma}_0
  \left( \frac{s_\mathrm{eq}}{s_\mathrm{y}} \right)^{1/m}
\end{equation}
where the yield stress is given by
\begin{equation}
  s_\mathrm{y} = s_0 + h \, \varepsilon_\mathrm{p}
\end{equation}
and finally the hardening
\begin{equation}
  h = h_0 \left( 1 - \frac{s_\mathrm{y}}{s_\infty} \right)^a
\end{equation}

\subsection{Material parameters and phase contrast}

\begin{table}
\centering
\caption{Used materials parameters for the crystal plasticity and the isotropic plasticity model. Young's modulus $E = 210$ GPa.}
\vspace*{0.2eM}
\begin{tabular}{l||c|c|c|c|c|c|c|c|c|}
Model & $s_0 / E$ & $s_\infty / E$ & $h_0 / E$ & $\dot{\gamma}_0$ & $a$ & $m$ & $q_n$ & $\nu$ \\
\hline
\hline
Isotropic plasticity & 170 & 8000 &  70 & 0.01 & 1.5 & 0.1 & 1.4 & 0.3  \\
Crystal plasticity   & 450 & 8000 & 800 & 0.01 & 1.5 & 0.1 & -   & 0.3  \\
\end{tabular}
\label{tab:parameters}
\end{table}

\begin{figure}[tph]
  \centering
  \includegraphics[width=0.5\linewidth]{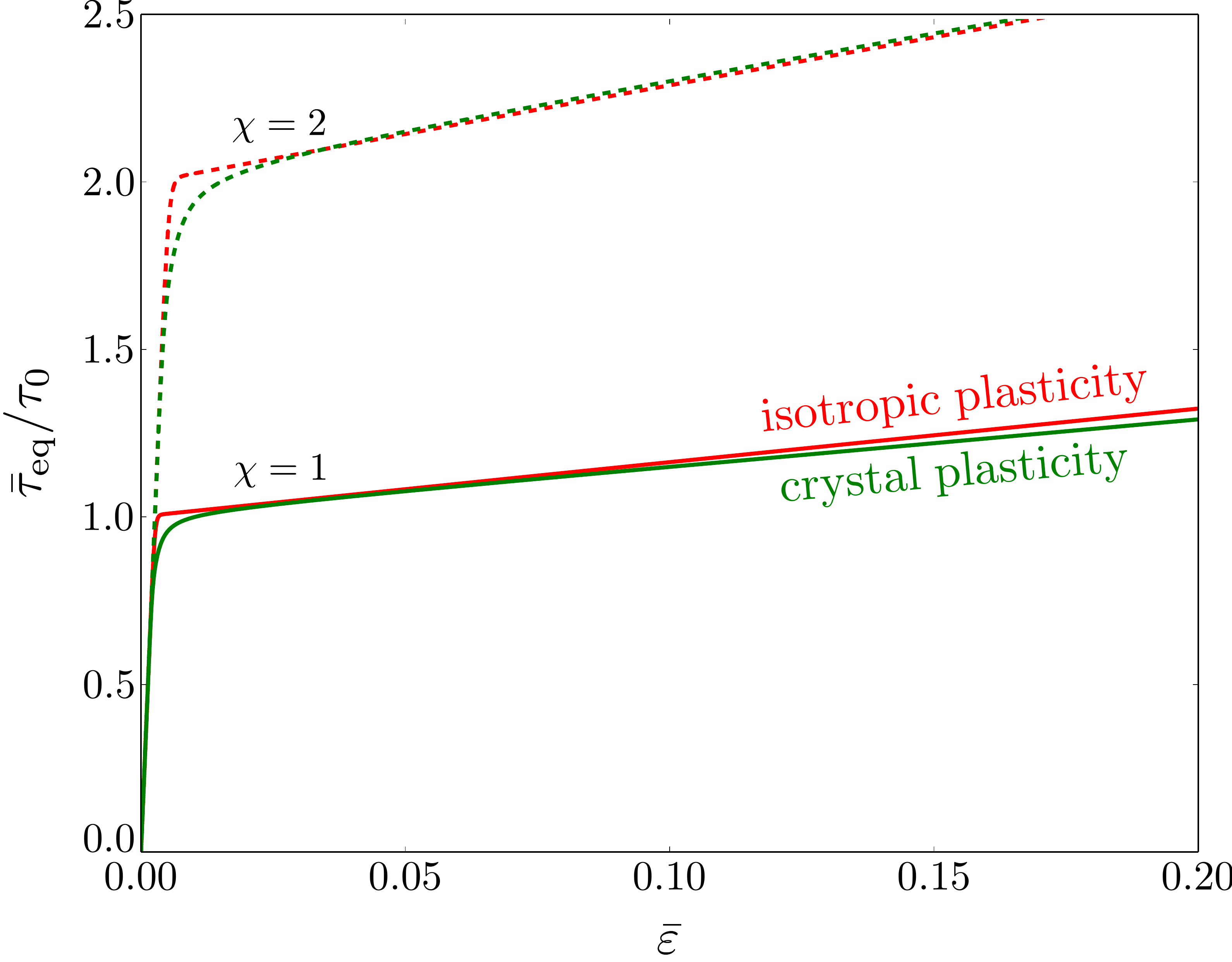}
  \caption{(Average) constitutive response of crystal plasticity (Section~\ref{sec:const:cp}) and isotropic plasticity (Section~\ref{sec:const:iso}), for two different hardening ratios: $\chi = 1$ and $\chi = 2$. N.B. no new parameter identification is performed for $\chi \neq 1$. }
  \label{fig:constitutive}
\end{figure}

The focus in this paper is on the plastic response, and hence the same (isotropic) elastic behavior is assumed for both models. To render the plastic response comparable, the parameters of the isotropic model are chosen representative for the considered class of materials \cite[e.g.][]{Vajragupta2012}. The soft phase is taken as a reference and the parameters of the hard phase are obtained by scaling them relative to the soft phase (see below). The constitutive response of the soft phase is plotted in Figure~\ref{fig:constitutive} using a solid red line, in terms of the equivalent Kirchhoff stress $\bar{\tau}_\mathrm{eq}$, as a function of the equivalent logarithmic strain $\bar{\varepsilon}$. Next, the parameters of the crystal plasticity model are fitted such that the response coincides with that of the isotropic model in an average sense (for a unit-cell that consists of many randomly oriented grains of the soft phase). The resulting average response is plotted in Figure~\ref{fig:constitutive} using a solid green line. Note that crystal plasticity predicts a more gradual transition from the elastic to the plastic regime, as the slip planes of the different grains become active at various stages of deformation. After yielding, the predicted response of the two models is comparable. All parameters are listed in Table~\ref{tab:parameters}.

Most parameters of the hard phase are the same as those of the soft phase, except for the yield stresses and hardening moduli -- $s_0$, $s_\infty$, and $h_0$ -- which are scaled by a factor $\chi = \tfrac{3}{2}$, $2$, $4$, and $8$. As an example, the behavior for $\chi = 2$ is shown in Figure~\ref{fig:constitutive} using dashed lines for both models. Note again that the plastic responses of both models closely resemble each other, except during initial yielding. A reference stress, $\tau_0$ is also introduced in Figure~\ref{fig:constitutive}. Its value coincides with the reference initial yield stress.

\section{Elasto-plastic response}
\label{sec:response}

\subsection{Macroscopic response}
\label{sec:macro}

\begin{figure}[tph]
  \centering
  \includegraphics[width=0.5\linewidth]{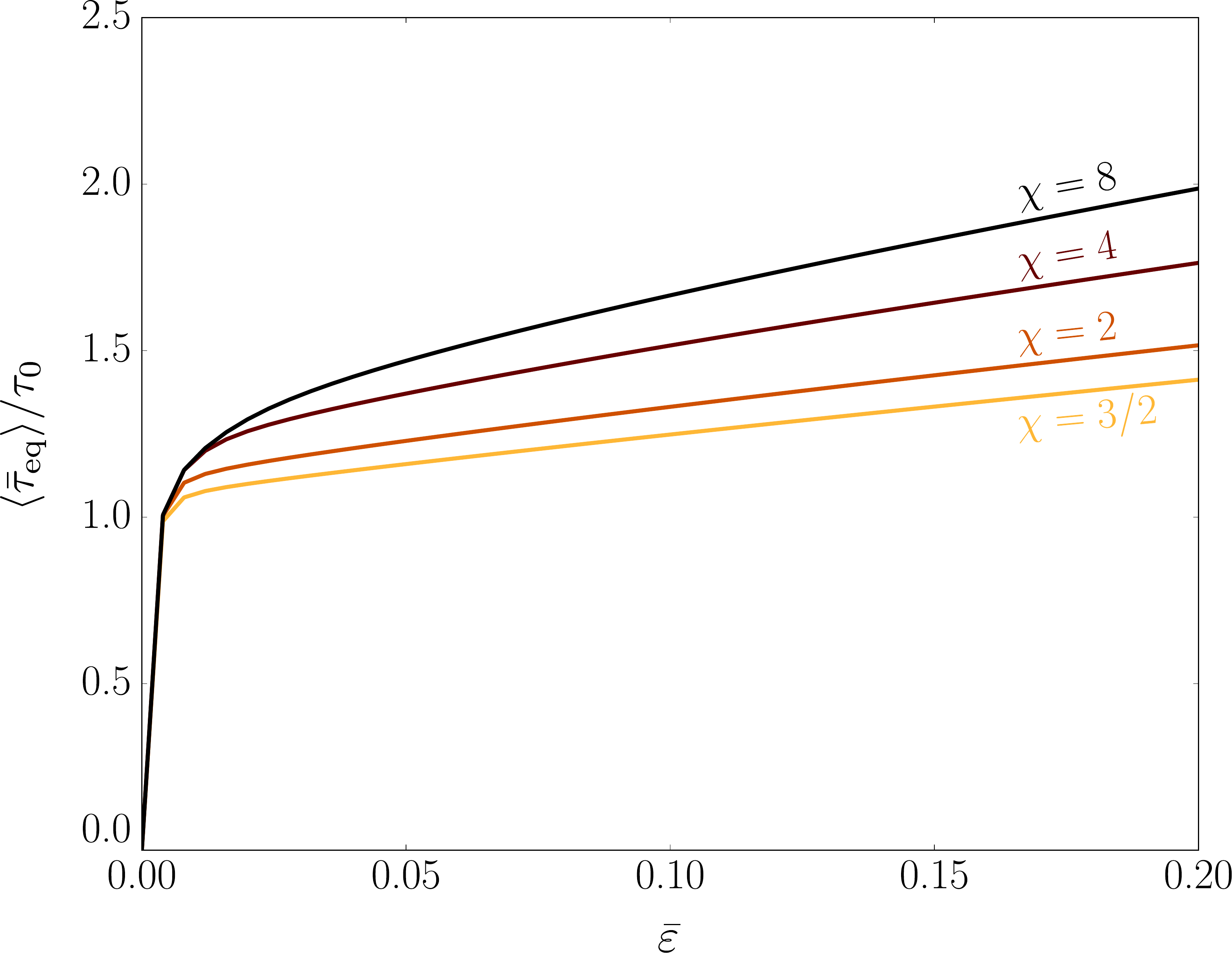}
  \caption{Normalized ensemble averaged macroscopic equivalent stress, $\langle \tau_\mathrm{eq} \rangle$, as a function of the applied equivalent strain, $\bar{\varepsilon}$, for different phase contrast, $\chi$. Both phases are modeled using crystal plasticity.}
  \label{fig:macroscopic_CP-CP_var-chi}
\end{figure}

The case for which both phases are modeled using crystal plasticity and for which the hard phase volume fraction $\varphi^\mathrm{hard} = 0.25$ is taken as reference. In Figure~\ref{fig:macroscopic_CP-CP_var-chi} the ensemble averaged  macroscopic equivalent Kirchhoff stress, $\langle \bar{\tau}_\mathrm{eq} \rangle$, normalized by the reference stress, $\tau_0$, is plotted as a function of the applied average strain, $\bar{\varepsilon}$, using different colors for different values of phase contrast, $\chi$. For low strains the response is more or less the same of all phase contrasts, while the hardening strongly increases with increasing $\chi$. This behavior results from the sequence of plastic events activating in the microstructure. First, the soft phase starts to yield, which is macroscopically observed as the onset of yielding. Then, due to the hardening of the soft phase, also the hard phase starts to yield. Since the yield stress (including hardening) of the hard phase increases with increasing $\chi$, more stress is required for the hard phase to yield, explaining the increased macroscopic hardening.

\begin{figure}[tph]
  \centering
  \includegraphics[width=1.\linewidth]{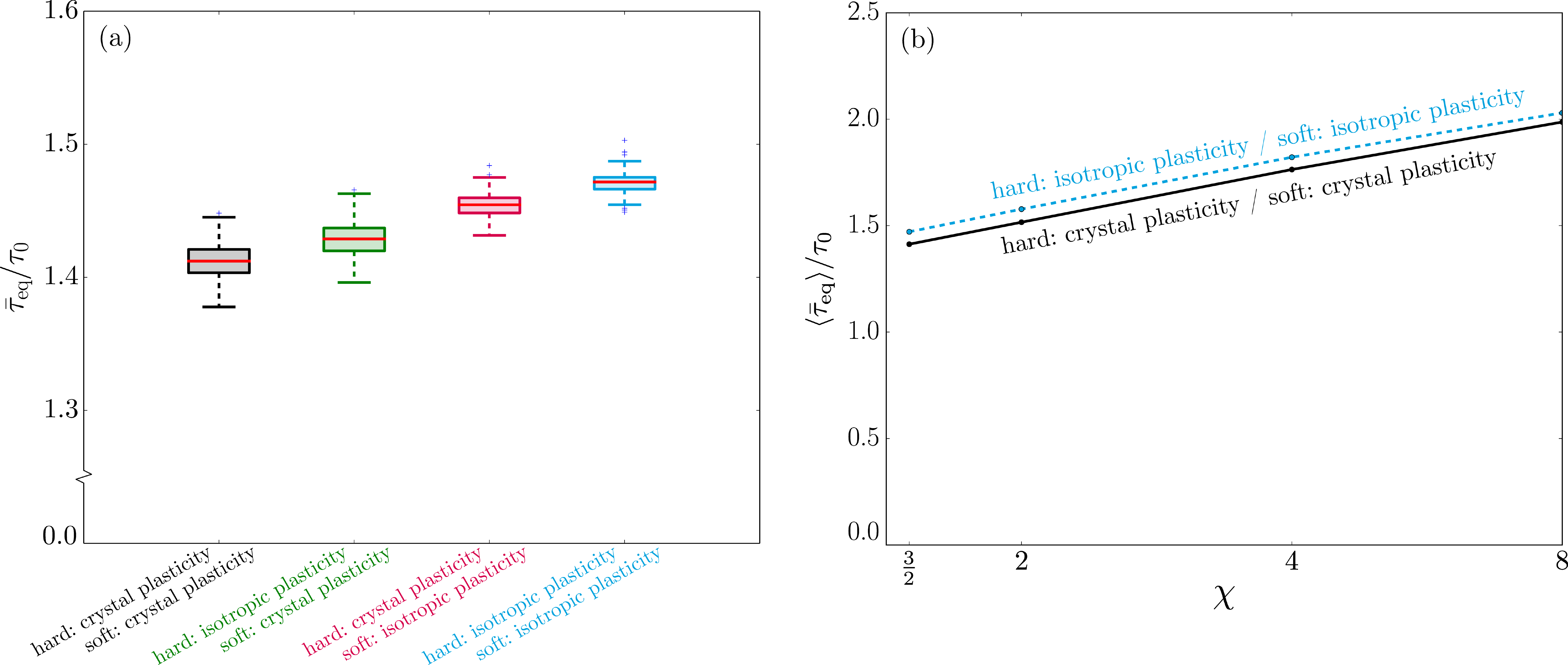}
  \caption{Effect of different constitutive assumptions for each of the phases on the macroscopic response, at an applied strain $\bar{\varepsilon} = 0.2$. (a) Box plots of the responses of all microstructures for all combinations of constitutive assumptions (for a phase contrast $\chi = 3/2$). (b) The ensemble averaged stress response in case that both phases are modeled using either crystal plasticity or isotropic plasticity, as a function of increasing phase contrast, $\chi$.}
  \label{fig:macroscopic_mat}
\end{figure}

The effect of using an isotropic model on the macroscopic stress response is shown in Figure~\ref{fig:macroscopic_mat} for an applied strain of $\bar{\varepsilon} = 0.2$. In Figure~\ref{fig:macroscopic_mat}(a) the microstructural model is simplified by applying the isotropic model for either one or both phases for a phase contrast $\chi = 3/2$. The ensemble averages and the scatter between the different microstructures are shown using box-plots. (i) The reference crystal plasticity model is shown in black (left). From left to right: (ii) the soft phase is assumed isotropic while the hard phase remains plastically anisotropic; (iii) the hard phase is assumed isotropic while the soft phase remains plastically anisotropic, and finally, (iv) both phases are assumed isotropic. In the extreme case with both phases isotropic, the macroscopic response is harder with respect to the reference case where both phases are plastically anisotropic. In addition, the scatter between the different microstructures is reduced. Furthermore, assuming the soft phase isotropic has a much bigger effect than assuming the hard phase isotropic (3\% instead of 1\% increase of $\bar{\tau}_\mathrm{eq}$). This can be understood from the fact that the soft phase takes most of the plastic deformation.

Figure~\ref{fig:macroscopic_mat}(b) shows the reference, crystal plasticity, model together with the strongest assumption where both phases are isotropic, as a function of phase contrast, $\chi$. The models slowly converge\footnote{This is also observed for an ensemble in which the hard phase volume fraction is fixed for each unit-cell (not shown here).} with increasing $\chi$: the isotropic model predicts a 4\% higher stress for $\chi = 3/2$ while this is reduced to 2\% for $\chi = 8$. This implies that if the contrast between the phases is large enough, the crystallographic mismatch between the grains is less important than the phase contrast, in terms of mechanical properties.

\subsection{Mesoscopic response}
\label{sec:response:micro}

\begin{figure}[tph]
  \centering
  \includegraphics[width=1.0\linewidth]{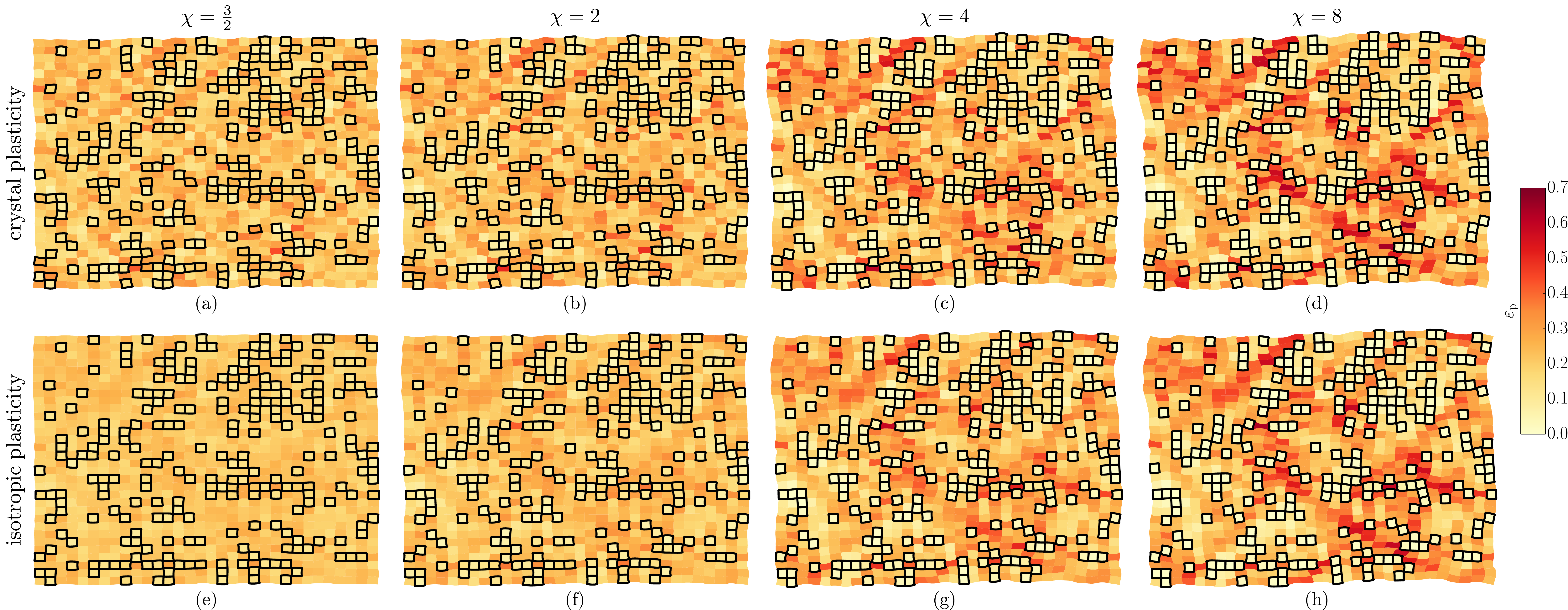}
  \caption{The grain-averaged equivalent plastic strain, $\varepsilon_\mathrm{p}$, for increasing phase contrast $\chi$ (left to right) and different constitutive models: both phases modeled using crystal plasticity (top) and both with isotropic plasticity (bottom); all at an applied strain of $\bar{\varepsilon} = 0.2$ and the same microstructure (include crystallography) of Figure~\ref{fig:typical_I_2D}. Grains of the hard phase are highlighted using a black outline.}
  \label{fig:typical_ep_var-chi}
\end{figure}

\begin{figure}[tph]
  \centering
  \includegraphics[width=0.5\linewidth]{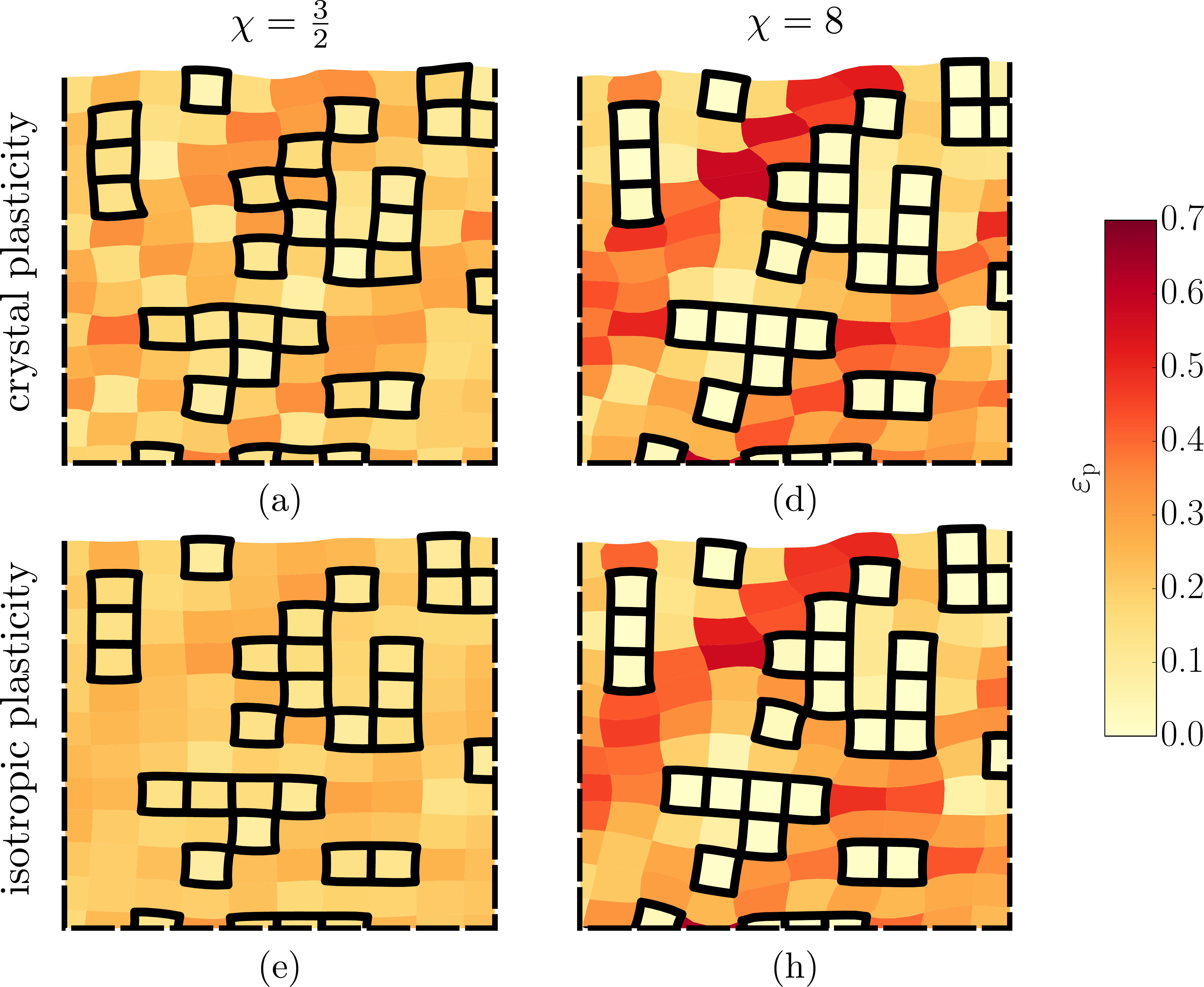}
  \caption{Zoom of Figure~\ref{fig:typical_ep_var-chi}: (a,e) for $\chi = 3/2$ and (d,h) for $\chi = 8$.}
  \label{fig:typical_ep_var-chi_zoom}
\end{figure}

The grain-averaged plastic response is shown in Figure~\ref{fig:typical_ep_var-chi} for a typical microstructure; all snapshots are taken at the same applied strain $\bar{\varepsilon} = 0.2$. The phase contrast $\chi$ increases from left to right. For the two extremes, $\chi = 3/2$ and $\chi = 8$, a magnified region is shown in Figure~\ref{fig:typical_ep_var-chi_zoom}. The reference model, where both phases are modeled using crystal plasticity, is shown in the top row, and the simplest approximation where both phases are isotropic is shown in the bottom row. In all cases, the plastic strain, $\varepsilon_\mathrm{p}$, is higher in the soft phase than in the hard phase. This partitioning is most pronounced for the highest phase contrast ($\chi = 8$, Figures~\ref{fig:typical_ep_var-chi}(d,h)) where almost no plastic strain is observed in the hard phase. The highest values of $\varepsilon_\mathrm{p}$ are found within bands in the soft phase oriented under $\pm 45$ degrees, coinciding with the shear directions; this deformation localization increases with $\chi$.

In comparing the two models (crystal vs.\ isotropic plasticity) only minor differences are observed for the highest phase contrast $\chi = 8$ (cf.\ Figures~\ref{fig:typical_ep_var-chi},\ref{fig:typical_ep_var-chi_zoom}(d,h)) both in the plastic strain, $\varepsilon_\mathrm{p}$, and the (periodic) displacements. Clear differences appear for lower $\chi$. In particular at the lowest phase contrast ($\chi = 3/2$), the displacements and the grain-averaged plastic strain are noticeably more homogeneous for the isotropic model (cf.\ Figures~\ref{fig:typical_ep_var-chi},\ref{fig:typical_ep_var-chi_zoom}(a,e)). For the crystal plasticity model, fluctuations of $\varepsilon_\mathrm{p}$ are observed from grain to grain within the soft phase, indicating that one grain is more favorably oriented to follow the applied deformation than the other. The local differences between the orientation of the slip systems of the different grains introduces a mechanical contrast that competes with the phase contrast.

\begin{figure}[tph]
  \centering
  \includegraphics[width=1.0\linewidth]{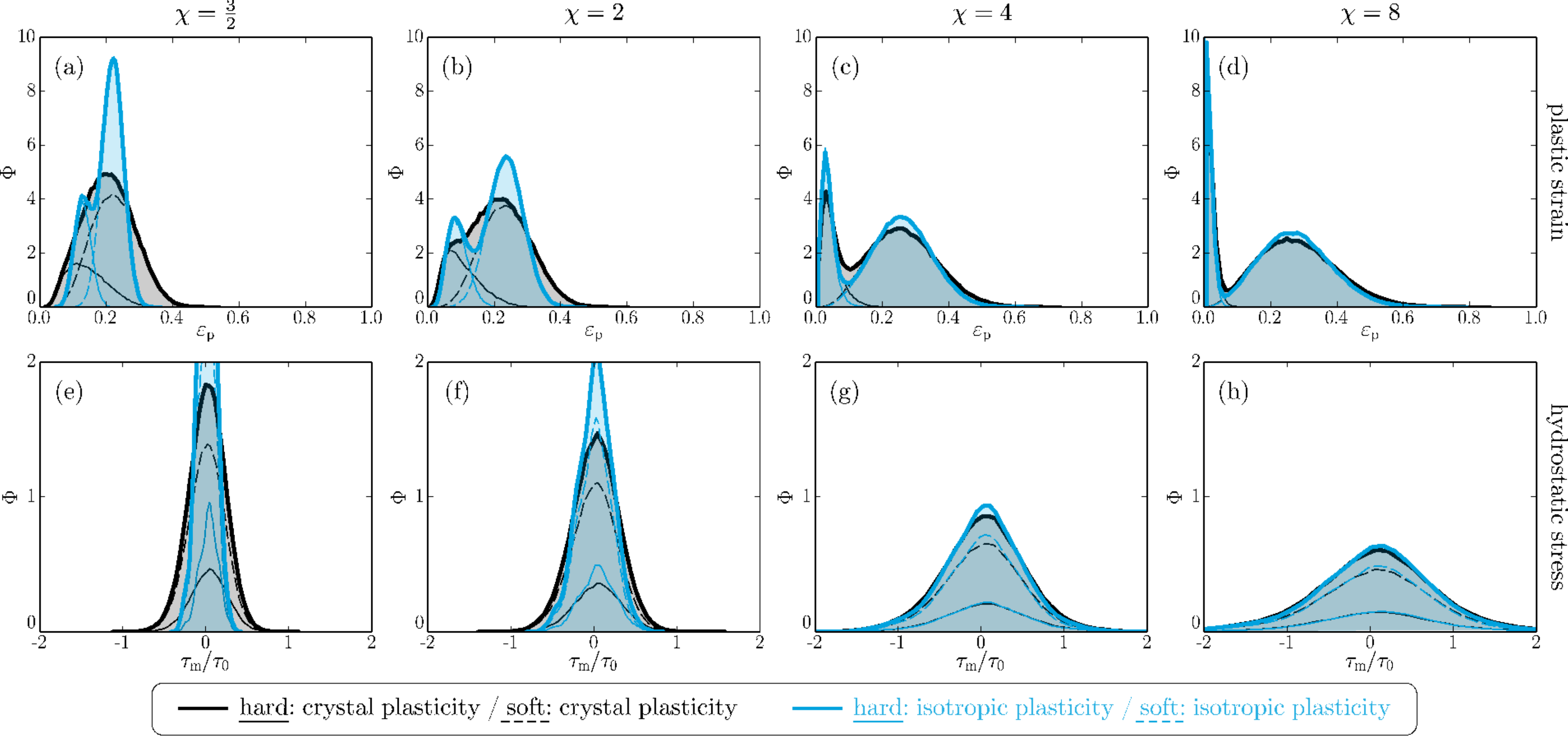}
  \caption{The ensemble probability density $\Phi$ of the grain-averaged plastic strain $\varepsilon_\mathrm{p}$ (top) and the grain-averaged hydrostatic stress $\tau_\mathrm{m}$ (bottom) at an applied strain $\bar{\varepsilon} = 0.2$. The phase contrast $\chi$ increases from left to right. The isolated hard phase (thin solid line) and soft phase (thin dashed line) are normalized with respect to their respective volume fractions. The different constitutive assumptions are plotted using a different color: black for crystal plasticity and blue for isotropic plasticity.}
  \label{fig:micro_CP-CP_vs_iso-iso_var-chi}
\end{figure}

To address the above observations systematically, the entire ensemble is considered. In Figure~\ref{fig:micro_CP-CP_vs_iso-iso_var-chi} the probability density $\Phi$ of the grain-averaged plastic strain $\varepsilon_\mathrm{p}$ (top) and hydrostatic stress $\tau_\mathrm{m}$ (bottom) are shown for increasing phase contrast $\chi$ (left to right). The plots include $\Phi$ of the ensemble of grains (thick solid line), as well as that of only the hard phase (thin solid line) and of only the soft phase (thin dashed line) -- for the reference crystal plasticity model (black) and the isotropic model (blue).

For the plastic strain (top) it is observed that the partitioning between the hard and the soft phase increases with increasing $\chi$: the soft phase accumulates more deformation. At the same time the scatter increases, implying that the plasticity is more localized. When the isotropic approximation (blue) is compared to the reference model (black), it is observed that the two tend to overlap with increasing $\chi$. This also holds for the plastic strain distribution within each phase\footnote{These observations, for the plastic strain and the hydrostatic stress, are also made for an ensemble in which the hard phase volume fraction is fixed for each unit-cell,.}.

For the hydrostatic stress (bottom), the amplitude is similar in the hard and the soft phase. Note that both phases have the same elastic response, justifying the comparable amplitudes. As the macroscopic deformation introduces no global hydrostatic stress, any non-zero hydrostatic stress is related to the inhomogeneity of the microstructure. Similarly as for the plastic strain, the hydrostatic stress in the isotropic approximation approaches that of the reference crystal plasticity model for increasing $\chi$.

\subsection{Influence of the hard phase volume fraction}
\label{sec:respose:vol}

\begin{figure}[tph]
  \centering
  \includegraphics[width=1.\linewidth]{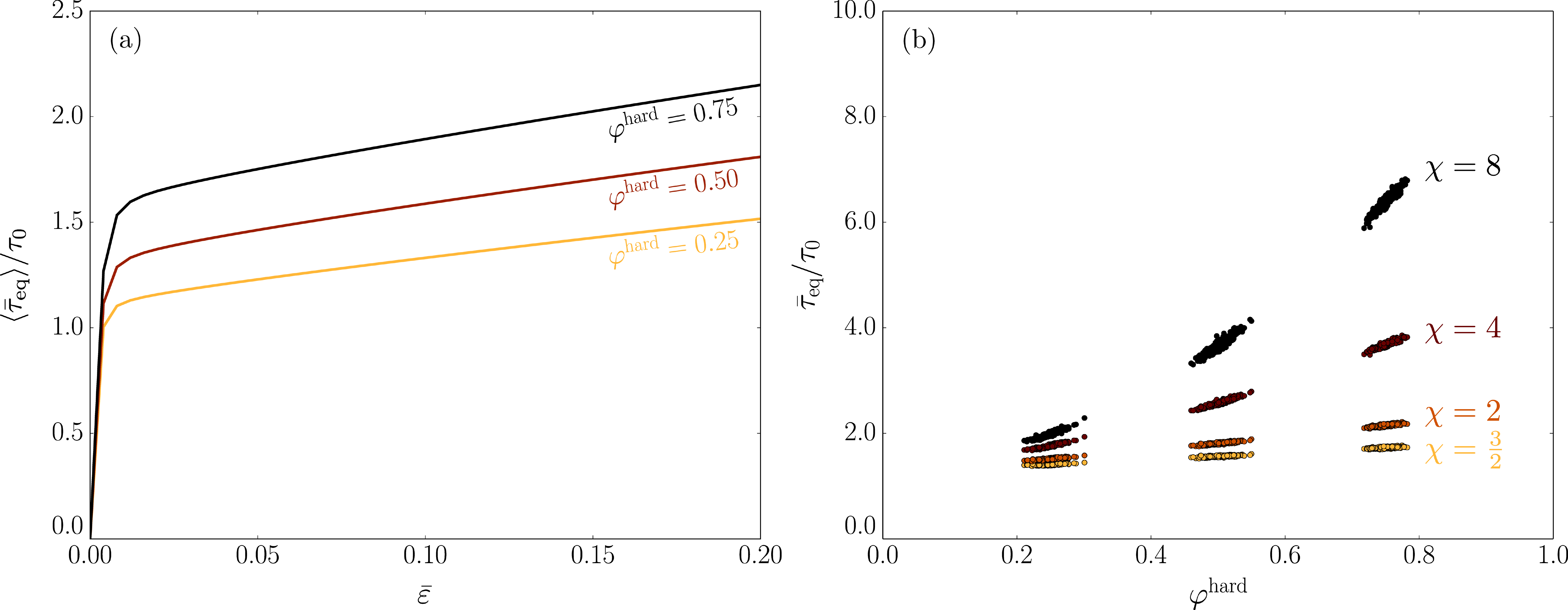}
  \caption{(a) The normalized ensemble averaged macroscopic equivalent stress, $\langle \tau_\mathrm{eq} \rangle$, as a function of the applied equivalent strain, $\bar{\varepsilon}$, for different hard volume fractions, $\varphi^\mathrm{hard}$; all for a phase contrast $\chi = 3/2$. (b) The response of the individual unit-cells at $\bar{\varepsilon} = 0.2$ for all parameter variations. The two phases are modeled using crystal plasticity.}
  \label{fig:macroscopic_CP-CP_var-phi}
\end{figure}

The analysis is extended to different ranges of hard phase volume fraction, i.e.\ $\varphi^\mathrm{hard} = 0.50$ and $0.75$. For the reference (crystal plasticity) model, the macroscopic response is shown in Figure~\ref{fig:macroscopic_CP-CP_var-phi}(a) using different colors for the different hard phase volume fractions; all ensembles have the same phase contrast $\chi = 3/2$. The initial yield stress and the hardening increase with increasing hard phase volume fraction, as to be expected from the rule of mixtures. The response at an applied strain $\bar{\varepsilon} = 0.2$ is shown for the full range of considered parameters in Figure~\ref{fig:macroscopic_CP-CP_var-phi}(b), where the macroscopic stress, $\tau_\mathrm{eq}$, is plotted as a function of the hard phase volume fraction, $\varphi^\mathrm{hard}$, using a marker for each individual unit-cell. The figure shows the strong combined influence of the hard phase volume fraction and phase contrast. The dependence on the volume fraction is clearly more pronounced for a larger phase contrast. For $\chi = 8$, $\bar{\tau}_\mathrm{eq}$ scales approximately with a factor $\chi \, \varphi^\mathrm{hard}$. For $\chi = 3/2$, it only scales with 50\% thereof. An empirical power-law scaling relation of the macroscopic stress relative to the volume fraction can thus be extracted from the responses:
\begin{equation}
  \frac{\bar{\tau}_\mathrm{eq}}{\tau_0}
  =
  1 + (\chi - 1)\, \varphi_\mathrm{hard}^{1.3}
\end{equation}
%

\begin{figure}[tph]
  \centering
  \includegraphics[width=1.0\linewidth]{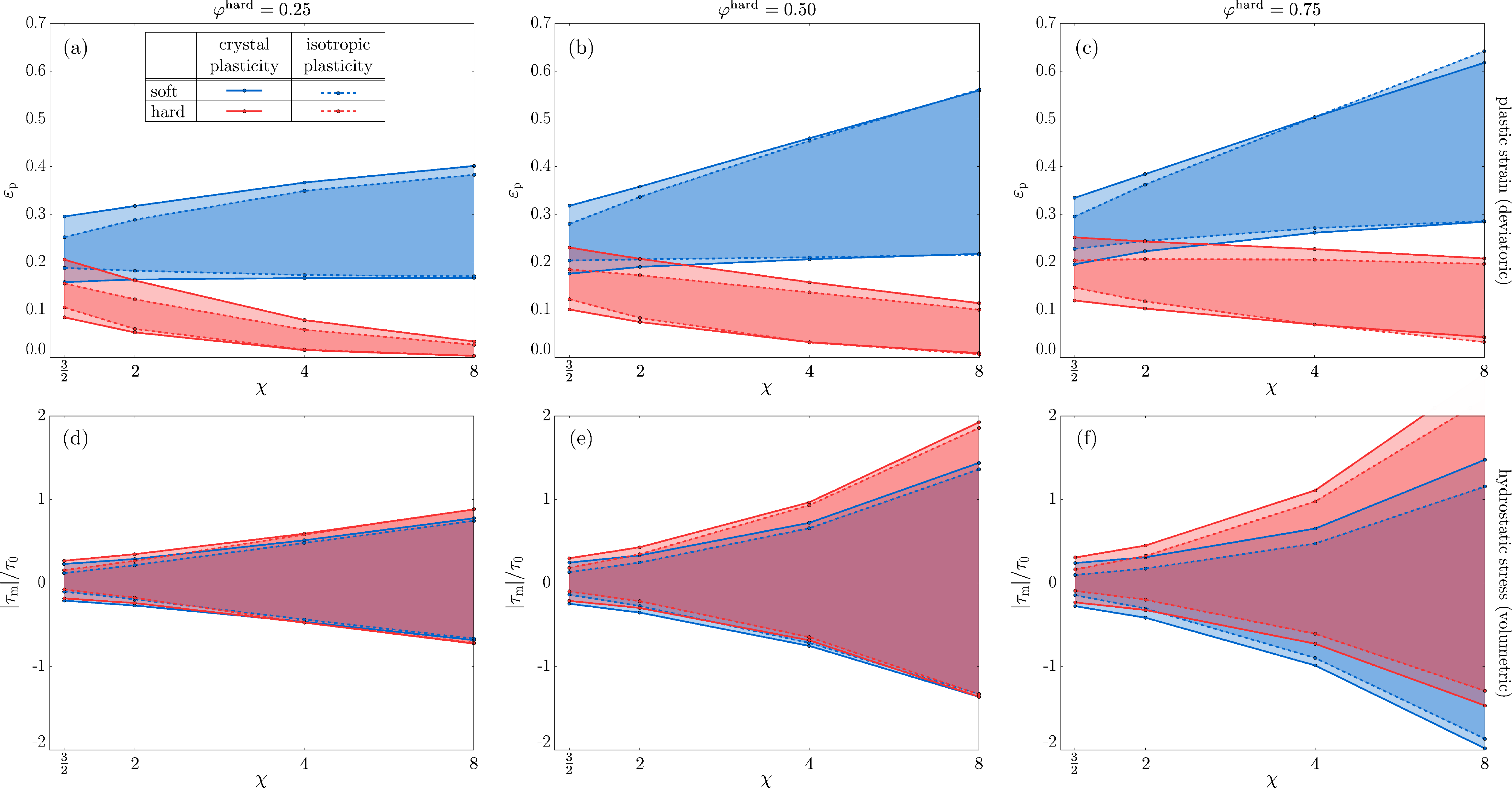}
  \caption{The distribution (average $\pm$ standard deviation of all grain-averaged responses in the ensemble) of the equivalent plastic strains $\varepsilon_\mathrm{p}$ (top) and hydrostatic stresses $\tau_\mathrm{m}$ (bottom), as a function of the phase contrast $\chi$. From left to right the volume fraction increases. Colors are used to discriminate between the hard phase (red) and soft phase (blue). The line-style corresponds to the constitutive model: both phases modeled using crystal plasticity (solid) and both phases with isotropic plasticity (dashed). Note that the averages are not shown to enhance interpretability but can be extracted as the middle between the lines indicating the averages $\pm$ the standard deviations. All figures are at $\bar{\varepsilon} = 0.2$.}
  \label{fig:micro_CP-CP_vs_iso-iso_std}
\end{figure}

The mesoscopic response is studied in a more coherent way by considering the distribution of the plastic strain and hydrostatic stress response of all grains in the ensembles, as a function of phase contrast $\chi$. Figure~\ref{fig:micro_CP-CP_vs_iso-iso_std} shows the distribution (confidence intervals) per phase: soft (blue) and hard (red). The solid lines correspond to the crystal plasticity model while the dashed lines correspond to the isotropic model. Figures~\ref{fig:micro_CP-CP_vs_iso-iso_std}(a--c) show the equivalent plastic strain, $\varepsilon_\mathrm{p}$, and Figures~\ref{fig:micro_CP-CP_vs_iso-iso_std}(d--f) show the hydrostatic stress, $\tau_\mathrm{m}$. From left to right the volume fraction increases. In each case the difference between the two models reduces with increasing phase contrast $\chi$, except for $\varphi^\mathrm{hard} = 0.75$. For $\varphi^\mathrm{hard} = 0.25$, the range of plastic strain is a factor of $2.2$ (soft phase) and $2.4$ (hard phase) more narrow for the isotropic model than for the crystal plasticity model for the lowest phase contrast, $\chi = 3/2$. In the other extreme, $\chi = 8$, these ratios reduce to $1.1$ and $1.3$, respectively. This is consistently observed also for the hydrostatic stress where for $\chi = 3/2$ the ratio is $2.0$ for the soft phase and $1.9$ for the hard phase, while they are reduced to $1.1$ and $1.05$ for $\chi = 8$. This trend is confirmed for $\varphi^\mathrm{hard} = 0.5$. For the highest hard phase volume fraction, $\varphi^\mathrm{hard} = 0.75$, however the range of plastic strains in the soft phase appears to be overestimated by the isotropic model for $\chi = 8$ (see Figure~\ref{fig:micro_CP-CP_vs_iso-iso_std}(c)) with a factor $1.1$ for the soft phase and $1.01$ for the hard phase, whereas the range of hydrostatic stress is underestimated by the isotropic model in both phases (with a factor of $1.2$ for the soft phase and $1.3$ for the hard phase, see Figure~\ref{fig:micro_CP-CP_vs_iso-iso_std}(f)). The mismatch originates from the fact that at this high hard phase volume fraction the hard phase starts to form a network across the unit-cell. In that case, the hard phase is forced to significantly deform plastically also at the highest $\chi$. The soft phase follows the hard phase, through the strong constraint acting on it.

\subsection{Effect of planar assumption}
\label{sec:response:3D}

To assess the consequences of assuming a two-dimensional microstructure, the microstructure is extended to three dimensions, again for the reference case for which the hard phase volume fraction $\varphi^\mathrm{hard} = 0.25$. Once again the grain-averaged responses predicted by the crystal plasticity model are compared to that of the isotropic model for different phase contrasts $\chi$.

An ensemble of 150 unit-cells comprising $12 \times 12 \times 12$ grains is used, for which the total number of grains is approximately the same as above. Each grain is discretized using a single 20 node tri-quadratic brick element, numerically integrated using eight Gauss points. Note that this discretization is coarse but functional.

\begin{figure}[tph]
  \centering
  \includegraphics[width=1.0\linewidth]{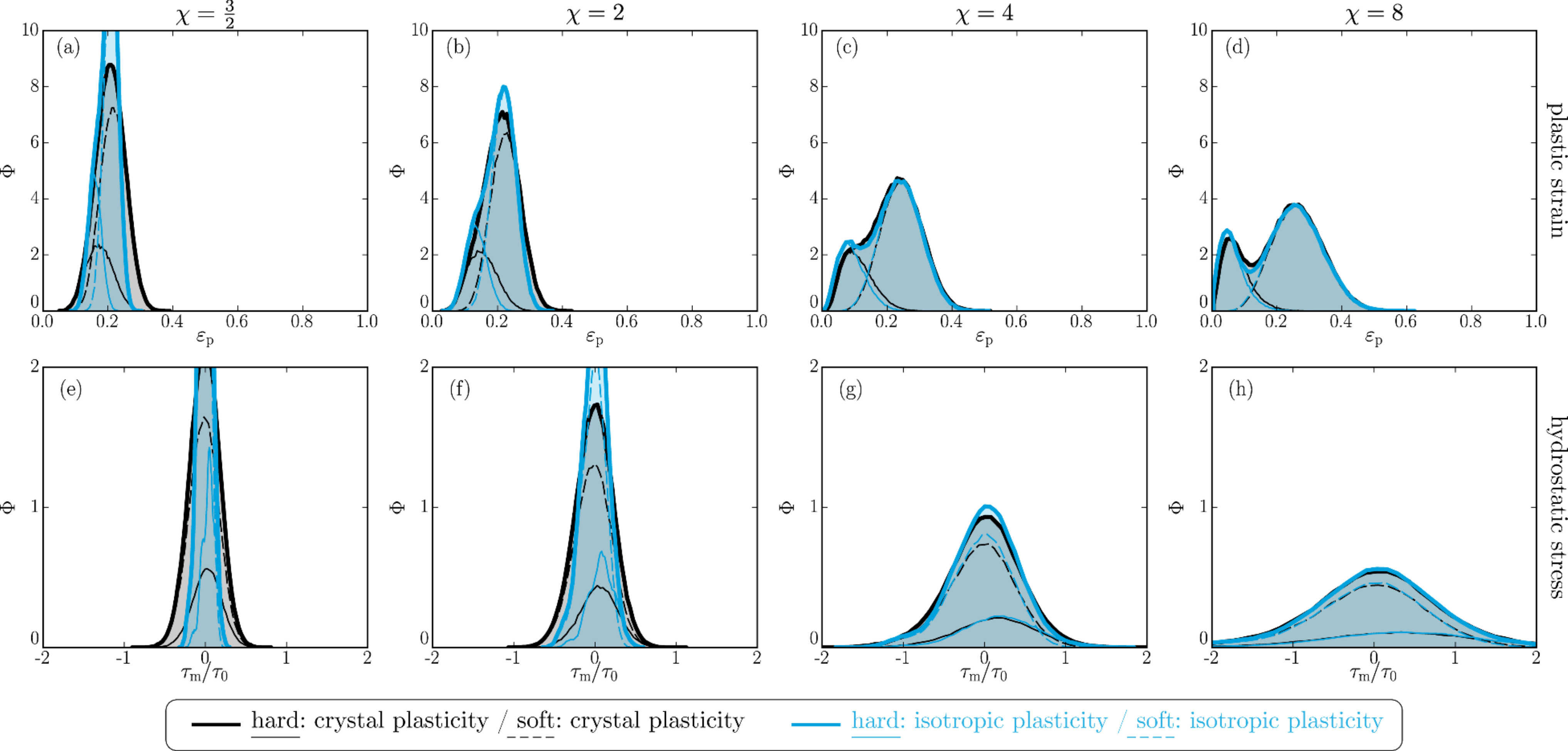}
  \caption{The ensemble probability density $\Phi$ of the grain-averaged plastic strain $\varepsilon_\mathrm{p}$ (top) and the grain-averaged hydrostatic stress $\tau_\mathrm{m}$ (bottom) for the ensemble of three-dimensional microstructures; cf.\ Figure~\ref{fig:micro_CP-CP_vs_iso-iso_var-chi}.}
  \label{fig:3D_pdf-ep-taum}
\end{figure}

The results are shown in Figure~\ref{fig:3D_pdf-ep-taum} using the probability density $\Phi$ of the plastic strain $\varepsilon_\mathrm{p}$ (top) and hydrostatic stress $\tau_\mathrm{m}$ (bottom), for increasing phase contrast $\chi$ (left to right). These results confirm the observations above. For the phase contrast $\chi = 4$ and $8$ the grain-averaged responses are accurately predicted using the isotropic model. For $\chi = 3/2$ and $2$ the isotropic model predicts more homogeneous grain-averaged responses than the crystal plasticity model. This indicates that only for the higher phase contrasts the effect of the crystallographic mismatch of the grains is overwhelmed by the phase contrast.

Compared to the 2-D results, the strain partitioning between the hard and the soft phase is less extreme for the 3-D microstructures (cf.\ Figure~\ref{fig:3D_pdf-ep-taum} with~\ref{fig:micro_CP-CP_vs_iso-iso_var-chi}). In the latter case, the sub-surface microstructure constrains the deformation. As a consequence, the hard phase is forced to deform more and consequently the soft phase deforms less. Macroscopically this results in a harder response of the actual 3-D microstructure compared with the 2-D approximation. These effects were investigated in more detail elsewhere \citep{DeGeus2016a}, however using the isotropic model only.

\subsection{Discussion}
\label{sec:response:discussion}

The results of the present study are consistent with earlier results in the literature. Both Needleman and Tvergaard \cite{Needleman1993} and McHugh et al.\ \cite{McHugh1993} using rigid reinforcement particles observed minor differences for an isotropic plasticity compared with a crystal plasticity model. This was experimentally confirmed by Nugent et al.\ \cite{Nugent2000}. This case corresponds best to a phase contrast of $\chi = 8$ in the present study, for which the same observations are revealed. If the reinforcement phase is no longer rigid, Choi et al.\ \cite{Choi2013} observed distinct differences between the isotropic and the reference crystal plasticity model. Their parameters compare with a phase contrast of $\chi = 3/2$ in the present study for which indeed clear differences appeared. Based on the results presented here, the apparently contradictory observations in the literature stem from the fact that each considers a clearly different regime for the phase contrast.

It may thus be concluded that if the mechanical contrast $\chi$ between the two phases is sufficiently high, the effect of the orientation discontinuity of the slip systems between individual grains is smaller or even negligible -- both in the volumetric and the deviatoric response. If the phase contrast is low, it competes with the contrast introduced by the orientation mismatch of the different crystals. Consequently, using isotropic plasticity is particularly meaningful for high phase contrasts. It can be used with an adequate quantitative approximation for $\chi \geq 4$. This conclusion also holds for different hard phase volume fractions, where the hard phase may no longer be considered as a reinforcement phase.

\begin{figure}[tph]
  \centering
  \includegraphics[width=0.65\linewidth]{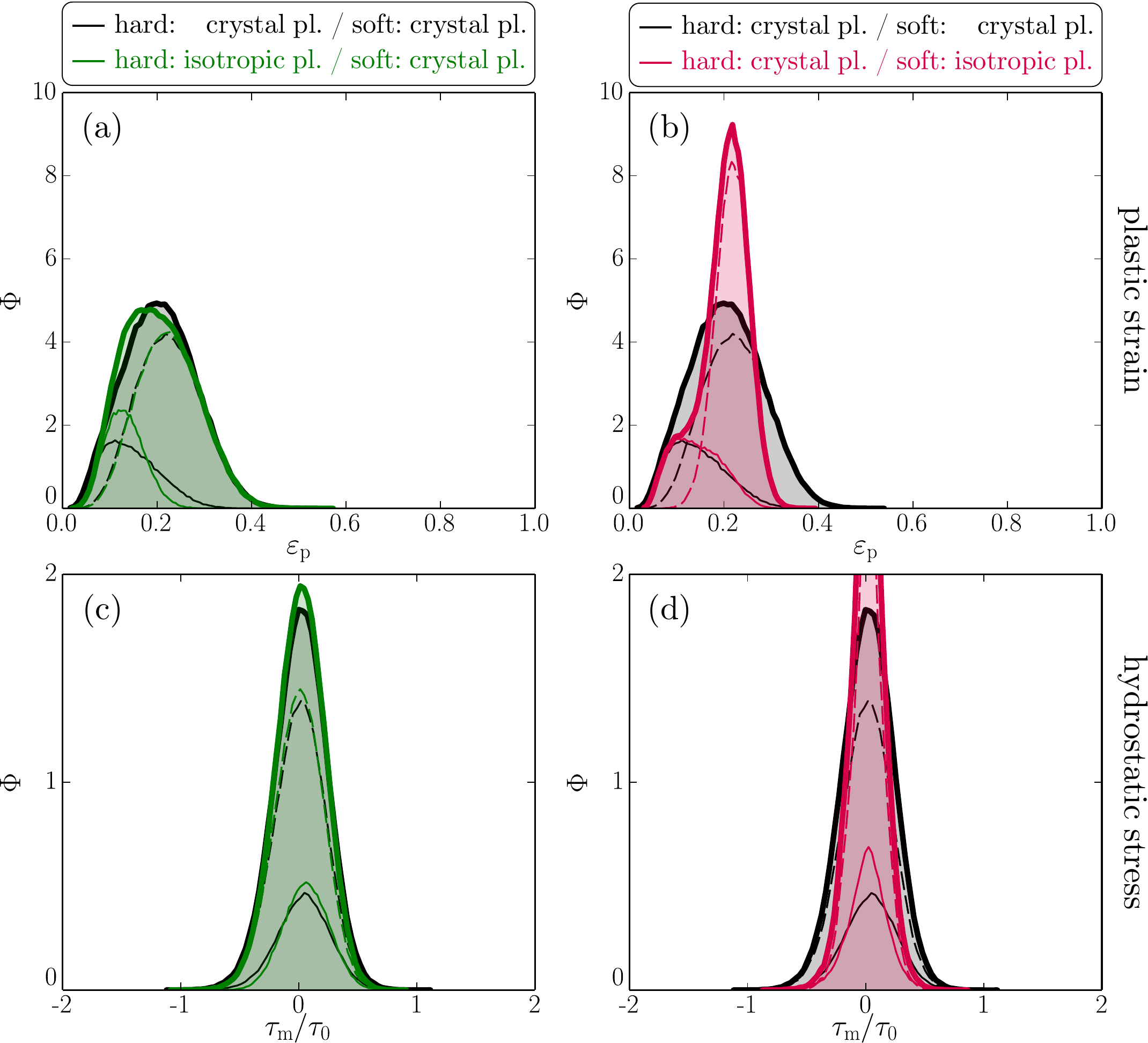}
  \caption{The probability density $\Phi$ of the plastic strain $\varepsilon_\mathrm{p}$ and hydrostatic stress $\tau_\mathrm{m}$ for the case where only one of the phases is isotropic: (a,c) the hard phase and (b,d) the soft phase. The phase contrast $\chi = 3/2$, the hard phase volume fraction $\varphi^\mathrm{hard} = 0.25$, and the applied strain $\bar{\varepsilon} = 0.2$.}
  \label{fig:micro_material-approx}
\end{figure}

For the phase contrast $\chi = 3/2$, where the crystallography plays an important role, it remains questionable whether the anisotropy of both phases has equal influence on the mechanical response. To this end, the reference ensemble is compared with two approximations in which only one of the two phases is assumed isotropic. The result is shown in Figure~\ref{fig:micro_material-approx} in terms of the ensemble probability density $\Phi$ of the equivalent plastic strain $\varepsilon_\mathrm{p}$ (top) and the hydrostatic stress $\tau_\mathrm{m}$ (bottom). When the hard phase is assumed isotropic (left), the plastic strain in the hard phase becomes more homogeneous; the plastic strain in the soft phase and the hydrostatic stress in both phases are not affected. On the other hand, if the soft phase is assumed isotropic only the plastic strain in the hard phase is unaffected. Both the plastic strain in the soft phase as well as the hydrostatic stress in both phases are significantly different. This observation is particularly interesting in the context of advanced high strength steels (the properties of which are in this range) as different researchers have confirmed that martensite, which acts as reinforcement phase, may be highly anisotropic in the plastic domain \cite{Maresca2014,Mine2013,Steinbrunner1988}.

\section{Ductile damage}
\label{sec:damage}

\subsection{Introduction}

The analysis is finally extended to ductile fracture initiation whereby emphasis is placed on two-phase steels where the hard phase acts as a reinforcement of the softer ductile matrix. Consequently, the hard phase volume fraction is fixed to $\varphi^\mathrm{hard} = 0.25$. The soft matrix is assumed to fail by ductile fracture while the hard phase is assumed not to fracture. Thereby we restrict ourselves to fracture initiation, for which it is assumed that fracture initiation sites do not interact. Consequently a damage indicator is used to signal fracture initiation. This indicator is not coupled to the constitutive response, i.e.\ no softening is considered. Though informative, this is of course a highly idealized analysis.

\subsection{Damage indicator}

Many different criteria exist for ductile fracture initiation, but they are all dependent on a non-linear combination of the plastic strain and hydrostatic stress \citep[e.g.][]{McClintock1968,Rice1969,Johnson1985}. Here, the Johnson-Cook damage indicator \cite{Johnson1985} is used to indicate the onset of ductile fracture in the soft phase. This indicator, $D$, signals fracture initiation once $D \geq 1$, but does not weaken the material response. It is defined as follows:
\begin{equation}
  D =
  \int_0^t
  \frac{ \dot{\varepsilon}_\mathrm{p} } { \varepsilon_\mathrm{c} ( \eta ) }
  \; \mathrm{d} \tau
\end{equation}
where $\dot{\varepsilon}_\mathrm{p}$ is the equivalent plastic strain rate and $\varepsilon_\mathrm{c}$ is a critical strain which depends on the local stress triaxiality $\eta$ in the following way
\begin{equation}
  \varepsilon_\mathrm{c} = A \exp ( - B \eta ) + \varepsilon_\mathrm{pc}
\end{equation}
wherein $A$, $B$, and the critical plastic strain $\varepsilon_\mathrm{pc}$ are material parameters. The stress triaxiality is the ratio of hydrostatic and equivalent (shear) Kirchhoff stress in the grain:
\begin{equation}
  \eta = \frac{ \tau_\mathrm{m} }{ \tau_\mathrm{eq} }
\end{equation}
The parameters are chosen representative for the considered class of materials, based on the literature \cite{Vajragupta2012}:
\begin{equation}
  A = 0.2 \qquad B = 1.7 \qquad \varepsilon_\mathrm{pc} = 0.1
\end{equation}
Note that the exact choice of parameters is not essential for the purpose of this study, however choosing them in a correct range leads to a more comprehensive interpretation of the value of $D$.

\subsection{Microscopic response}

\begin{figure}[tph]
  \centering
  \includegraphics[width=1.0\linewidth]{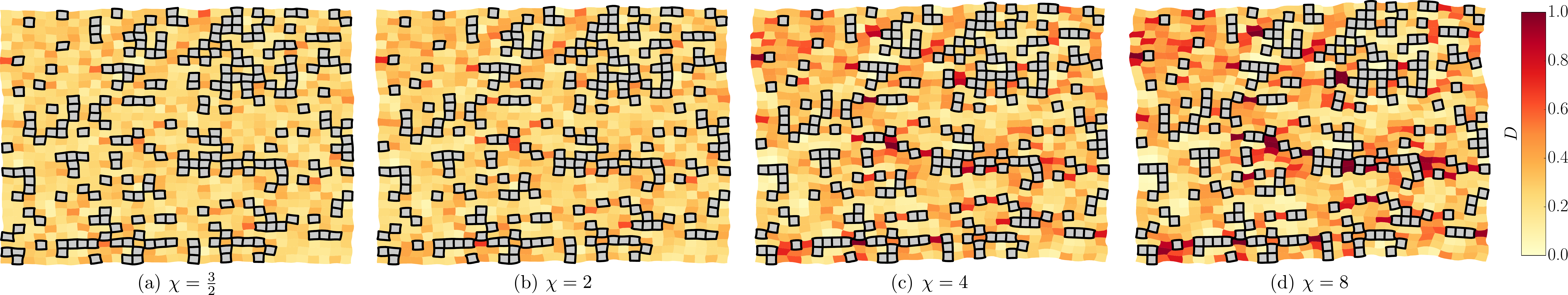}
  \caption{The grain-averaged damage indicator $D$ in the soft phase for different phase contrasts $\chi$. The microstructure and applied deformation is the same as in Figure~\ref{fig:typical_I_2D}, both phases are modeled using crystal plasticity.}
  \label{fig:typical_D_var-chi}
\end{figure}

For the microstructure presented earlier in Figures~\ref{fig:typical_I_2D} and \ref{fig:typical_ep_var-chi}, the damage indicator $D$ is shown in Figure~\ref{fig:typical_D_var-chi} for different values of the phase contrast $\chi$ (at the same applied strain $\bar{\varepsilon} = 0.2$). For an increasing phase contrast $\chi$, the damage indicator also increases, indicating that a higher phase contrast promotes damage. Furthermore, the damage is more localized for higher $\chi$. It is particularly high near the interfaces of the hard and the soft phase (discussed in more detail in \cite{DeGeus2015a}). In a large part of the shear bands that are observed in Figure~\ref{fig:typical_ep_var-chi}, $D$ is significantly lower than one.

\begin{figure}[tph]
  \centering
  \includegraphics[width=1.0\linewidth]{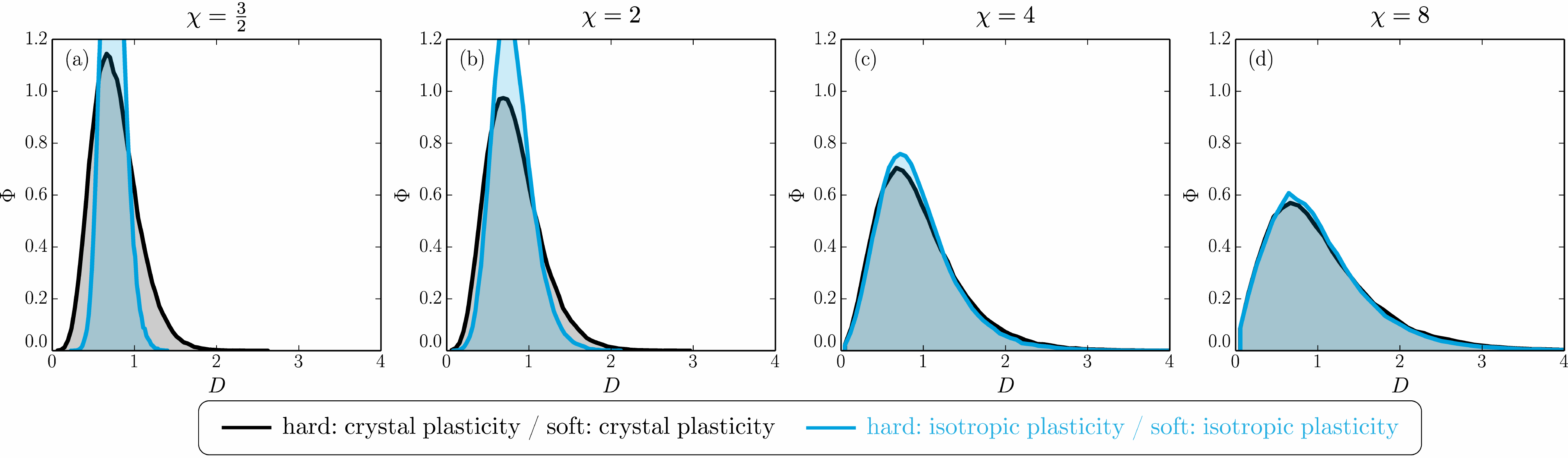}
  \caption{The probability density $\Phi$ as a function of the damage indicator $D$ in the soft phase, for different values of the phase contrast $\chi$. The applied strain $\bar{\varepsilon} = 0.2$.}
  \label{fig:D_CP-CP_vs_iso-iso}
\end{figure}

The importance of the plastic anisotropy is assessed by comparing the assumption of plastic isotropy to the reference crystal plasticity model. Figure~\ref{fig:D_CP-CP_vs_iso-iso} shows the probability density function $\Phi$ of the damage $D$ for different levels of phase contrast $\chi$. The reference model is shown in black, while the isotropic model is shown in blue. For the lowest phase contrast, $\chi = 3/2$, there is a significant difference between the damage predicted by the two models, whereby the damage is considerably higher in the crystal plasticity model in $12\%$ of the grains. The value of $D$ predicted by the two models converges with increasing $\chi$ (it is wrongly predicted in only 1\% of the grains for $\chi = 8$). Based on these results, it may be concluded that for $\chi \geq 4$ there is no pronounced difference between isotropic and crystal plasticity, while for $\chi \leq 2$ the difference is significant.

\subsection{Effect of the morphology: damage hot-spot}

Since the damage is dominated by the presence of the two phases, the question arises how their spatial distribution contributes to this, and whether this contribution is different when also the crystallography significantly contributes to damage. To quantify this, the average phase arrangement is characterized as a function of the position relative to the fracture initiation sites. Following De Geus et al.\ \cite{DeGeus2015a}, the probability of finding the hard phase is calculated as a function of the position relative to the fracture initiation site. As only two phases are considered, the probability of finding the soft phase is implicitly contained in this result. If at a certain relative position the probability of hard phase is higher than the hard phase volume fraction, $\varphi^\mathrm{hard}$, having hard phase at this relative position thus promotes fracture initiation. Likewise, if the probability is lower than $\varphi^\mathrm{hard}$, having soft phase at that relative position promotes fracture initiation.

This method is discussed based on a single unit-cell. The extension to the ensemble average is trivial. A phase indicator $\mathcal{I}$ is introduced that describes the phase distribution as follows:
\begin{equation}
  \mathcal{I}(i,j) = \begin{cases}
    1 \quad\mathrm{for}\; (i,j) \in \mathrm{hard} \\
    0 \quad\mathrm{for}\; (i,j) \in \mathrm{soft}
  \end{cases}
\end{equation}
where $(i,j)$ is the position, here coinciding with the \textit{matrix-index} of each grain. Similarly, fracture initiation is indicated by $\mathcal{D}(i,j)$ which is one if the damage indicator $D(i,j) \geq 1$ and zero elsewhere. To correlate the arrangement of phases to the fracture initiation, the average value of $\mathcal{I}$ is calculated at a certain position $(\Delta i, \Delta j)$ relative to the fracture initiation sites, where $\mathcal{D} = 1$, as follows:
\begin{equation}
  \mathcal{I}_\mathcal{D} (\Delta i, \Delta j) =
  \frac{
    \sum_{ij} \;
    \mathcal{D}(i,j) \;
    \mathcal{I} ( i+\Delta i, j+\Delta j)
  }{
    \sum_{ij} \;
    \mathcal{D} (i,j) \hfill
  }
\end{equation}
where $(i,j)$ loops over all grains, taking the cell's periodicity into account. The resulting damage hot-spot $\mathcal{I}_\mathcal{D}$ thus characterizes the probability of finding the hard phase at different positions $(\Delta i, \Delta j)$ relative to fracture initiation. The ensemble average $\langle \mathcal{I}_\mathcal{D} \rangle$ trivially follows by looping over all grains in the ensemble.

\begin{figure}[tph]
  \centering
  \includegraphics[width=1.0\linewidth]{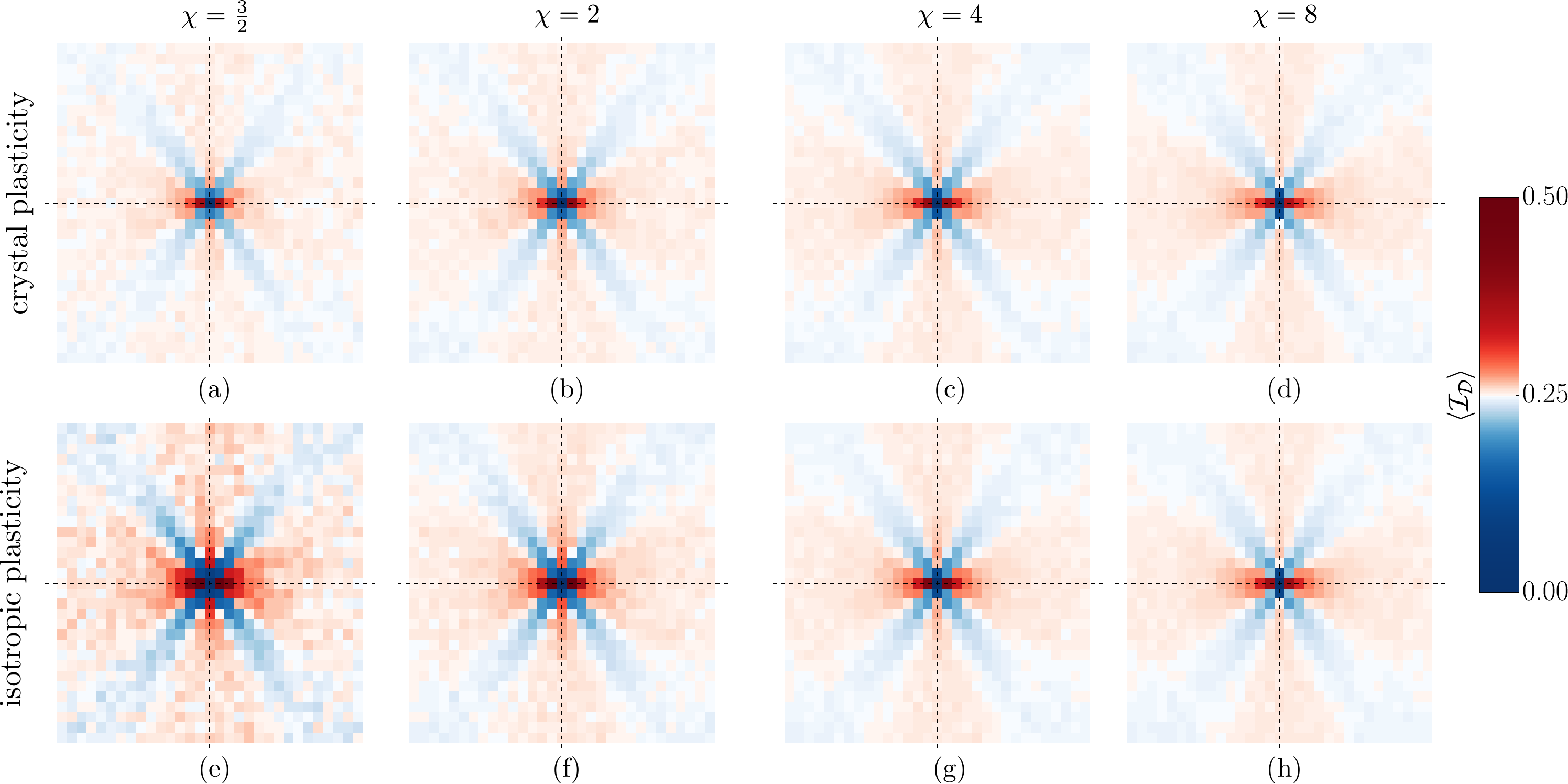}
  \caption{The average phase indicator around fracture initiation, $\mathcal{I}_\mathcal{D}$, for different phase contrasts $\chi$, for a hard phase volume fraction $\varphi^\mathrm{hard} = 0.25$ and applied strain $\bar{\varepsilon} = 0.2$. The rows correspond to different constitutive choices: both phases modeled using crystal plasticity (top) or using isotropic plasticity (bottom). The colormap is chosen such that the neutral color (white) coincides with the hard phase volume fraction. The colors may be interpreted as: hard phase (red) and soft phase (blue).}
  \label{fig:hotspot_CP-CP_vs_iso-iso}
\end{figure}

The result is shown in Figure~\ref{fig:hotspot_CP-CP_vs_iso-iso} for different phase contrasts $\chi$, whereby the response to the crystal plasticity model is shown on the top row and the response to the isotropic plasticity model on the bottom row. The colormap has been chosen to enhance the interpretation, whereby red indicates an elevated probability of the hard phase and blue to that of the soft phase, the neutral color (white) coincides with the hard phase volume fraction $\varphi^\mathrm{hard}$ indicating no preferential phase at that position relative to damage. The results show that fracture initiates where regions of the hard phase are aligned with the tensile axis and interrupted by regions of the soft phase aligned with the shear direction at $\pm 45$ degree angles (see \cite{DeGeus2015a}, and reference therein, for a detailed mechanical analysis). Qualitatively the arrangement of phases around fracture initiation sites appears to be independent of the phase contrast and of the plastic (an)isotropy. Even though the influence of crystallography on the value of the damage, $D$, may not be neglected for a low phase contrast; the spatial locations where fracture initiates are not affected by it.

\section{Conclusions}
\label{sec:conclusion}

This paper studies the role that the plastic anisotropy, as introduced by the crystallography, plays on the mechanical response at the meso-scale level of a two-phase metal. The grain-averaged responses as predicted by a crystal plasticity model are compared to those predicted by an isotropic formulation; using idealized two-dimensional microstructures. The following observations have been made.

\begin{itemize}
  \item The plastic anisotropy is less important than the phase contrast when the ratio of the yield stresses between the phases is larger than four. Below a ratio of two the influence of the orientation mismatch of the grains is significant. This has been assessed through the plastic strain and the hydrostatic stress. This confirms observations in the literature where only one of the regimes was analyzed independently \citep{Needleman1993,McHugh1993,Choi2013}.
  \item These observations also hold for the initiation of ductile fracture of the comparatively soft phase.
  \item The spatial arrangement of the phases plays a crucial role for the initiation of ductile fracture in the soft matrix phase \citep[e.g.][]{Brockenbrough1991,McHugh1993}. The analysis showed that this effect dominates over the role of crystallography, even in the regime where its quantitative influence on the mechanical response may not be neglected.
\end{itemize}

\section*{Acknowledgments}

This research was carried out under project number M22.2.11424 in the framework of the research program of the Materials innovation institute M2i (\href{http://www.m2i.nl}{www.m2i.nl}).

\bibliography{library}

\end{document}